# BUNCH-BY-BUNCH DETECTION OF COHERENT TRANSVERSE MODES FROM DIGITIZED SINGLE-BPM SIGNALS IN THE TEVATRON*

G. Stancari[#†], A. Valishev, and A. Semenov, FNAL, Batavia, IL 60510, U.S.A.

*Abstract*

A system was developed for bunch-by-bunch detection of transverse proton and antiproton coherent oscillations based on the signal from a single beam-position monitor (BPM) located in a region of the ring with large amplitude functions. The signal is digitized over a large number of turns and Fourier-analyzed offline with a dedicated algorithm. To enhance the signal, the beam is excited with band-limited noise for about one second, and this was shown not to significantly affect the circulating beams even at high luminosity. The system is used to measure betatron tunes of individual bunches and to study beam-beam effects. In particular, it is one of the main diagnostic tools in an ongoing study of nonlinear beam-beam compensation studies with Gaussian electron lenses. We present the design and operation of this tool, together with results obtained with proton and antiproton bunches.

## INTRODUCTION

In the Tevatron, 36 proton bunches collide with 36 antiproton bunches at the center-of-momentum energy of 1.96 TeV. Each species is arranged in 3 trains of 12 bunches, circulating at a revolution frequency of 47.7 kHz. The bunch spacing within a train is 396 ns, corresponding to 21 53-MHz rf buckets. The bunch trains are separated by 2.6-μs abort gaps. The synchrotron frequency is around 30 Hz, or $7\times10^{-4}$ times the revolution frequency. The machine operates with betatron tunes near 20.58.

The betatron tunes and tune spreads of individual bunches are affected by the head-on and long-range beam-beam interaction. These phenomena are among the limiting factors of modern colliders. For optimal machine performance, knowledge of bunch-by-bunch tune distributions is crucial. Three systems are currently used in the Tevatron to measure incoherent tune distributions: the 21.4-MHz Schottky detectors, the 1.7-GHz Schottky detectors, and the direct diode detection base band tune (3D-BBQ). The latter two can be gated on single bunches.

Detection of transverse coherent modes can complement these three systems. The relationship between coherent oscillation frequencies and betatron tunes is indirect, and therefore a systematic uncertainty arises from the beam-beam model when extracting the lattice tunes and the beam-beam parameter. Still, the method is appealing because of its high frequency resolution.

*Work supported by the United States Department of Energy under contract No. DE-AC02-07CH11359.

[#]stancari@fnal.gov

[†]On leave from Istituto Nazionale di Fisica Nucleare, Sezione di Ferrara, Italy.

Another motivation for measuring coherent modes is our study of nonlinear beam-beam compensation with Gaussian electron lenses [1-3]. Due to their high brightness compared with protons, an improvement in antiproton lifetime from beam-beam compensation is not foreseen. The goal of this project is a proof-of-principle observation of tune shifts induced by the electron lens and the associated changes in tune spread and beam-beam parameter. This may benefit the Relativistic Heavy Ion Collider, where electron lenses are being built, and possibly the Large Hadron Collider. Also in the case of beam-beam compensation, the spectrum of coherent modes complements the observation of Schottky signals in terms of frequency resolution.

In this paper, we present the detection technique and the measured spectra of proton and antiproton bunches under various experimental conditions.

## TRANSVERSE COHERENT MODES

Transverse coherent modes carry information about lattice tunes and the beam-beam parameter [4-8]. In the simplest case, when two identical bunches collide head-on in one interaction region, two modes appear: a σ-mode at the lattice tune, where bunches oscillate transversely in phase, and a π-mode, separated from the σ-mode by a shift slightly larger than the beam-beam parameter, in which bunches are out of phase.

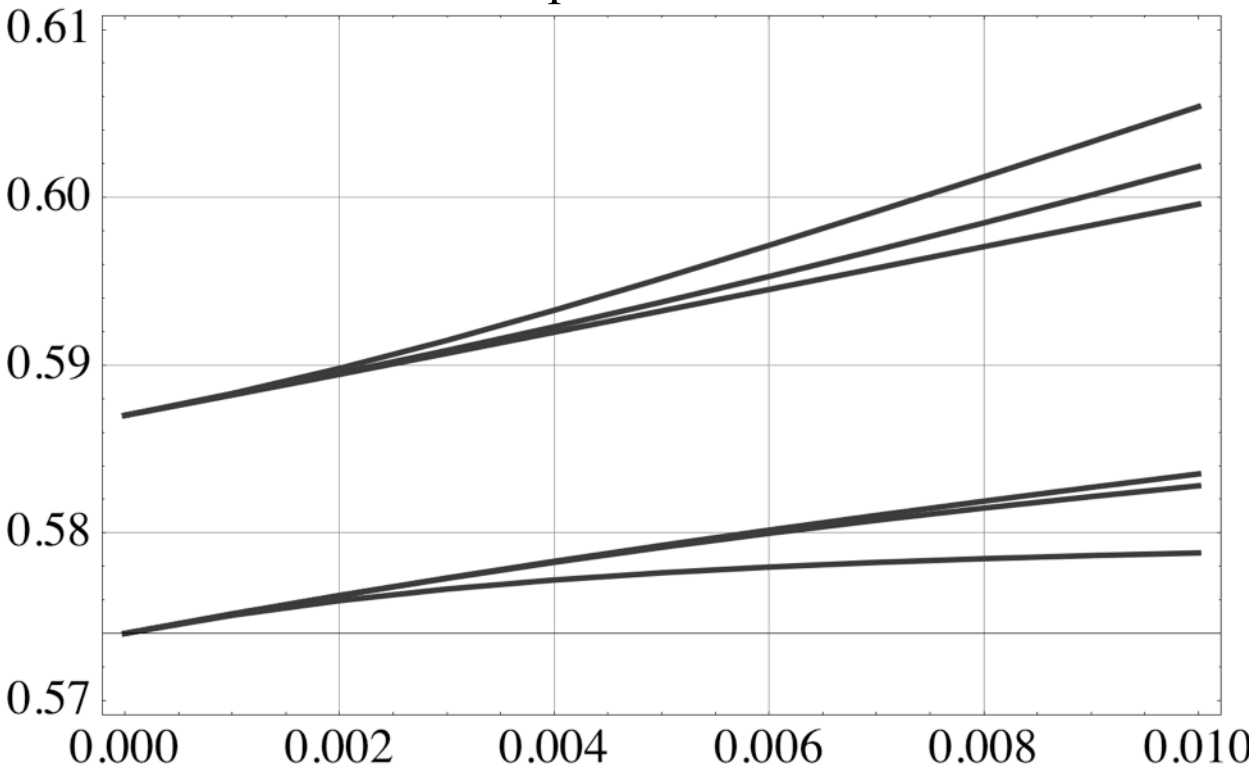

Figure 1: Sample calculation of coherent mode frequencies vs. beam-beam parameter per interaction point for 3×3 rigid round bunches with different lattice tunes.

In general, the frequency and number of these modes depend on the number of bunches and interaction regions, on the tune separation between the two beams, on beam sizes and relative intensities. Figure 1 shows how the mode frequencies evolve as a function of the beam-beam parameter, in the case of 3 rigid round bunches colliding with 3 bunches with different lattice tune at 2 head-on crossings. This simplified model captures the essential features of collisions at the Tevatron.

Coherent beam-beam modes were observed in several electron machines, including PETRA, LEP, and VEPP-2M [4,9,10]. Although their observation in hadron machines is made more difficult by the lack of strong damping mechanisms, they were seen both at the ISR and at RHIC [11,12].

## APPARATUS

The system (Figure 2) is based on the signal from a single vertical BPM (VB11) located near the B0 interaction point, in a region where the amplitude function at collisions is 900 m. It is a stripline pickup, with two plate outputs (A and B) for both protons and antiprotons. While observing the proton signal, the outputs can be split, sending half of the signal to the Tevatron BPM system and orbit stabilization and half to the present system. Antiproton signal are a factor three weaker and are usually not used for orbit feedback, so the splitter is not necessary and the full signal can be analyzed. Switching between proton and antiproton signals presently requires physically swapping cables.

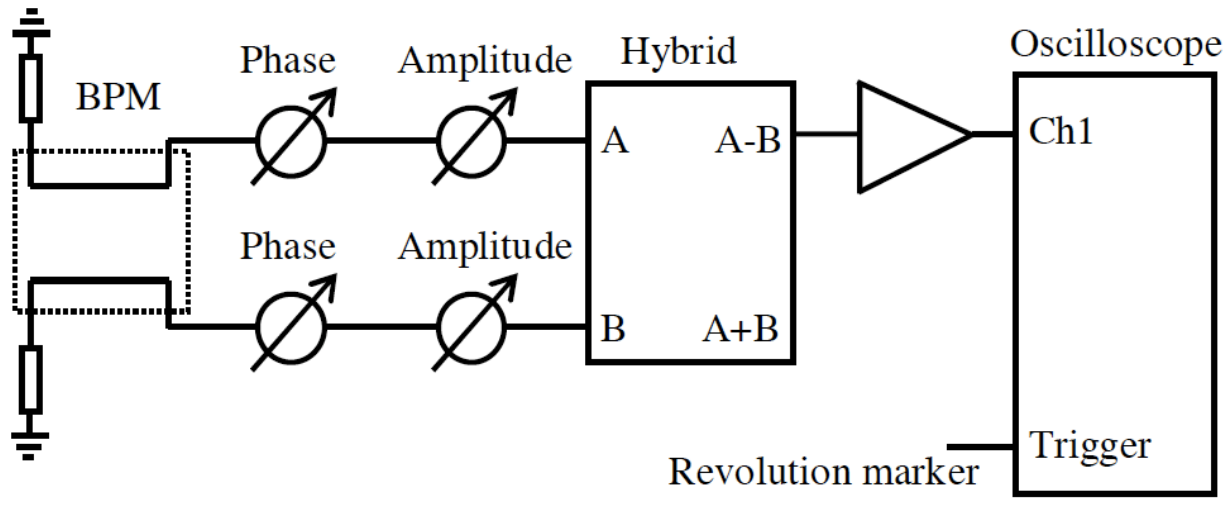

Figure 2: Schematic diagram of the apparatus.

Before processing, it is necessary to equalize the A and B signals to take advantage of the full dynamic range of the digitizer without saturating it. Equalization also reduces false transverse signals due to trigger jitter, as discussed below. The phase and attenuation of each signal is manually adjusted before feeding it to the hybrid by looking at the A–B output. If necessary, fine-tuning is done by displacing the beam with a small orbit bump. Orbits at collisions are stable over a time scale of weeks, and this adjustment does not need to be repeated often. To automate the task in the case of changing orbits and intensities (e.g., for observations at top energy between the squeeze and initiating collisions, or for observing both proton and antiproton bunches), a circuit board is being designed with self-calibrating gains and offsets.

The difference signal from the hybrid is amplified by 23 dB and sent to the digitizer. We use a 1-channel, 1-V full range, 10-bit digitizer (Agilent Acqiris series) with time-interleaved ADCs that can sample at 8 GS/s and store a maximum of 1024 MS or 125,000 segments (due to a firmware problem, only half of the memory could be used). The 47.7-kHz Tevatron revolution marker is used as trigger, so we will refer to 'segments' or 'turns' interchangeably. Typically, we sample at 8 GS/s (sample period of 125 ps), which corresponds to 150 slices for each 19 ns rf bucket. A C++ program running under Windows on the front-end computer controls the digitizer.

To enhance the signal, the beam is excited with a few watts of band-limited noise ('tickling') for about 1 s during the measurement, which simply consists of digitizer setup, tickler turn-on, acquisition start, tickler turn-off, and acquisition stop. The cycle takes a few seconds. The procedure is parasitical and it was shown not to affect the circulating beams even at high luminosity ($3.5\times10^{32}$ cm$^{-2}$ s$^{-1}$). When repeating the procedure several times continuously, the 21.4-MHz Schottky spectrum showed some activity, but no beam loss was observed.

## DATA ANALYSIS

Data analysis is performed offline using the multi-platform, open-source R package [13] running on the front-end computer or wherever is convenient. Data is written in binary format. It contains the raw ADC data together with trigger time stamps and delays of the first sample with respect to the trigger. Timing information has an accuracy of about 15 ps, and it is extremely important for the synchronization of samples.

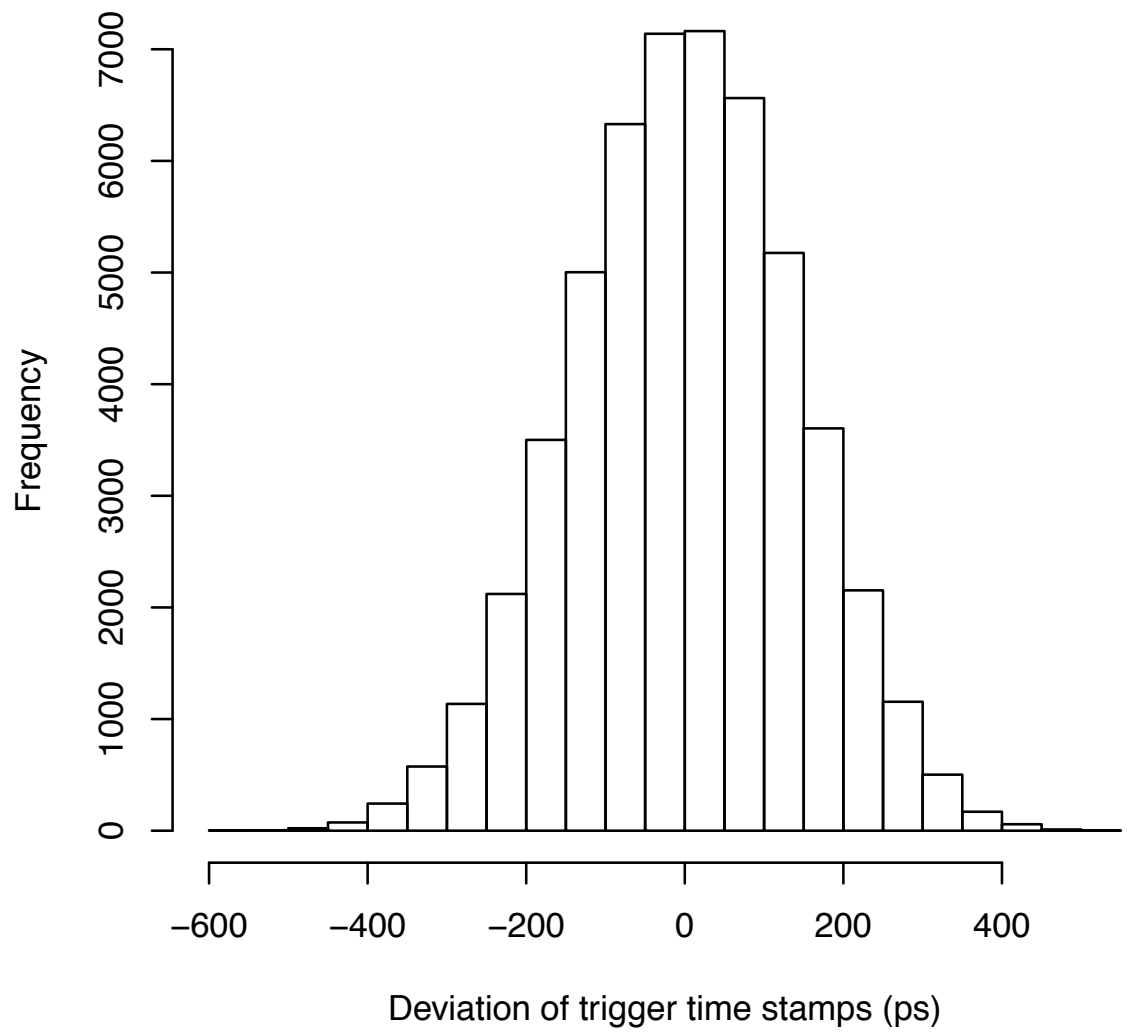

Figure 3: Distribution of the deviations of the trigger time stamp from the ideal trigger time for all 52,707 turns in a measurement.

From the distribution of differences between consecutive trigger time stamps, the average revolution frequency is calculated and, from it, the ideal trigger time stamps for each turn. The distribution of the difference between measured and ideal time stamps (Figure 3) is a measure of the jitter in the revolution marker, and its rms is typically 0.2 ns. The sum of trigger offset and sample delay is the correction by which each sample in a segment is to be shifted in time to be aligned with the other segments. For each turn and each bunch, the signal is interpolated with a natural spline and shifted according to the correction. The only side effect of this procedure is that a few slices at the edges become unusable, as they cannot be replaced with real data. The synchronization of turns is extremely important, as the jitter in trigger translates into a false transverse oscillation where the signal has a slope. If the BPM signals are not well balanced, jitter of even a fraction of a nanosecond can raise the noise floor by several dB and compromise the measurement.

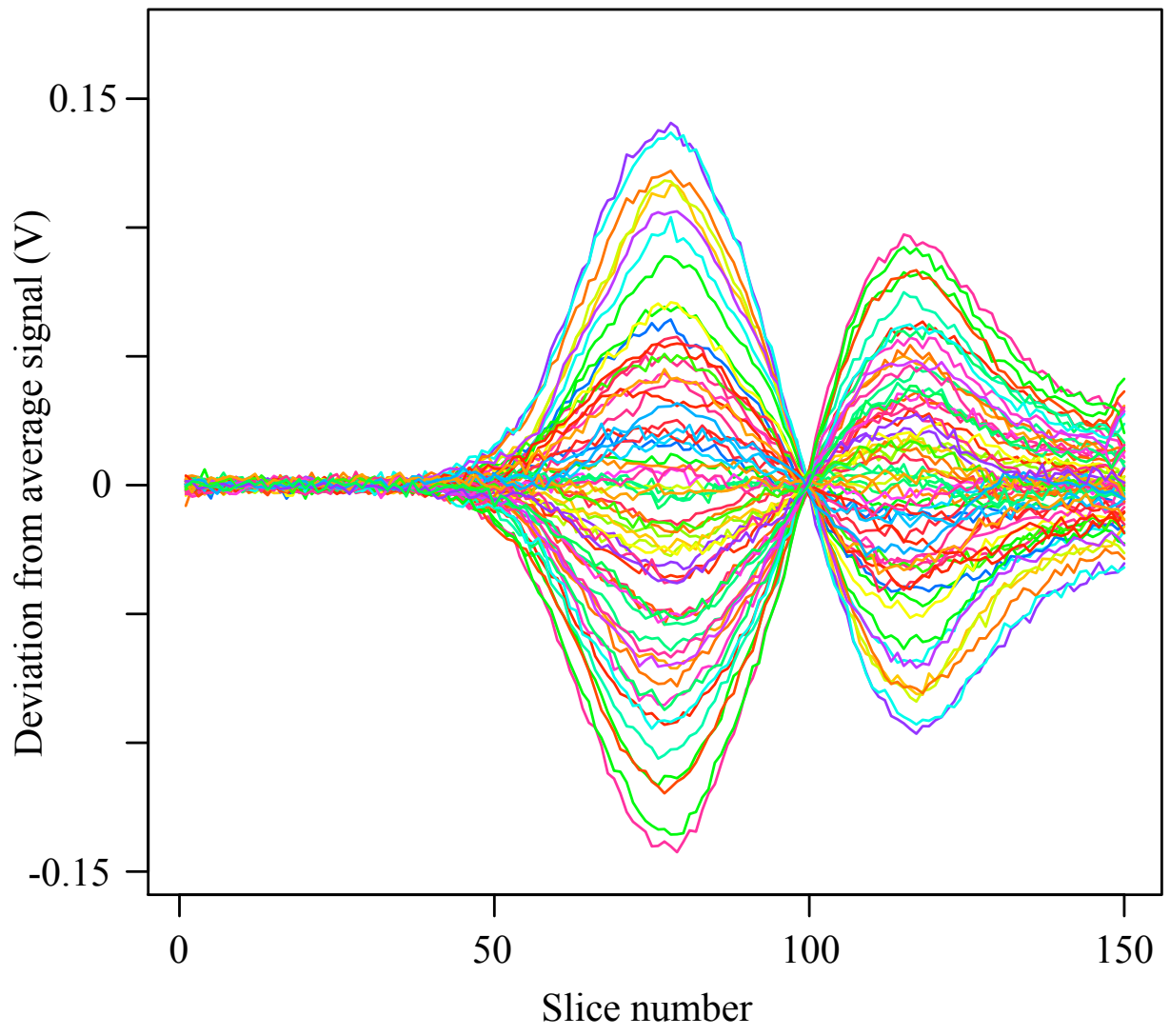

Figure 4: Example of digitized antiproton signal.

Figure 4 shows an example of digitized data for an antiproton bunch after synchronization and offset subtraction. Each slice corresponds to 125 ps. The full vertical scale represents a displacement of the beam centroid of about 20 µm. 64 randomly chosen turns are plotted out of the recorded 52,707.

For each bunch, the signal of each individual slice vs. turn number is Fourier transformed. The number of bins in the FFT vector is chosen according to the desired frequency resolution, $10^{-4}$ of the revolution frequency being a typical choice. The data is multiplied by a Slepian window of rank 2 to confine leakage to the adjacent bins and suppress it below $10^{-5}$ in farther bins. To reduce data loss from windowing, the FFT vectors are overlapped and averaged.

The noise level was estimated by observing the spectra without beam. They show several sharp lines. These are attributed to gain and offset differences between the time-interleaved ADCs themselves and to timing skew of their clocks [14-16]. The same spurious lines are also present in the Fourier spectrum of the time stamps, and this corroborates their attribution to digitizer noise.

To improve the signal-to-noise ratio, and to suppress backgrounds unrelated to the beam, such as the spurious lines from the digitizer, a set of signal slices (near the signal peaks) and a set of background slices (before the arrival of the beam) are defined. Power spectrums are extracted for both signal and background, and their ratio is calculated. The ratios are very clean, with some additional variance at the frequencies corresponding to the narrow noise spikes.

Data analysis takes about 50 s per bunch for 62,500 turns, its duration being dominated by the synchronization algorithm, which is interpreted within the R environment. Of course, it can be compiled, if necessary, and linked for faster response.

## RESULTS

In preparation for our studies on beam-beam effects, the system was tested on both proton and antiproton bunches under different experimental conditions.

Beam position variations are dominated by long-term drifts, which are compensated by the orbit-stabilization program, and by low-frequency jitter, related to ground motion and other sources [17-20]. This low-frequency jitter does not affect the measurements directly, but it reduces the available dynamic range. (A high-pass filter and more amplification are being considered to improve the system.)

At injection (150 GeV), while antiprotons are being loaded into the machine on a different orbit, the proton signal shows a single main peak at the lattice tune, even without tickling (Figure 5, top). At collisions (Figure 5, bottom), a complex pattern of beam-beam modes with synchrotron sidebands appears, showing some dependence on longitudinal position in the bunch.

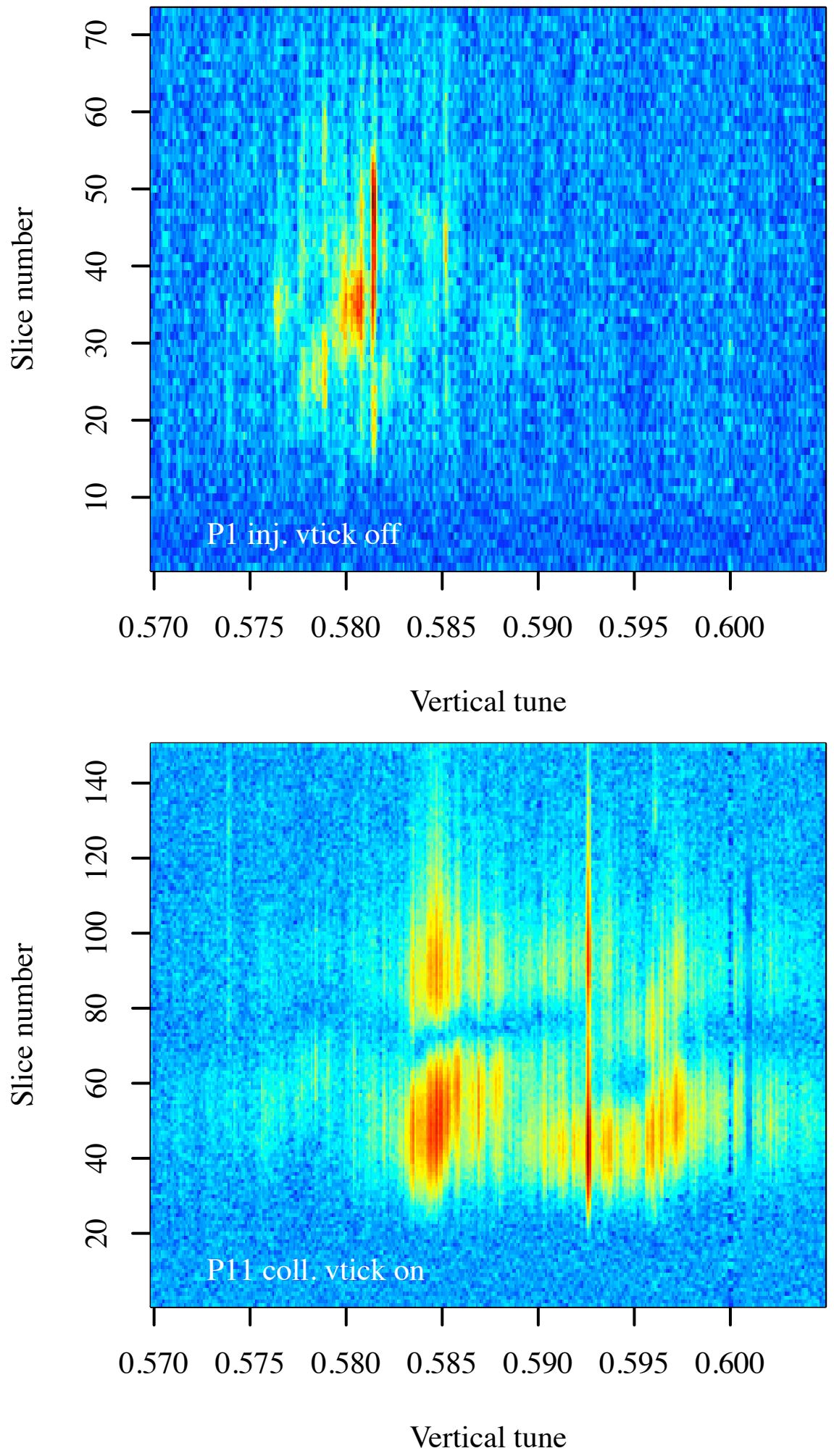

Figure 5: Comparison between the spectra of protons at injection (top) and after initiating collisions (bottom)

It was also verified that the mode frequencies respond to changes in tunes, and that the π-mode gets stronger as the tunes of the two beams are brought closer together.

The evolution of transverse coherent modes in the vertical plane over a complete store was studied with antiprotons (Figure 6). The lattice tune was intentionally changed between the last three measurements to keep the average Schottky tune constant. A decrease in mode separation can be observed, in agreement with the decrease of beam-beam forces. One can also appreciate the large ratio between signal level and background variance.

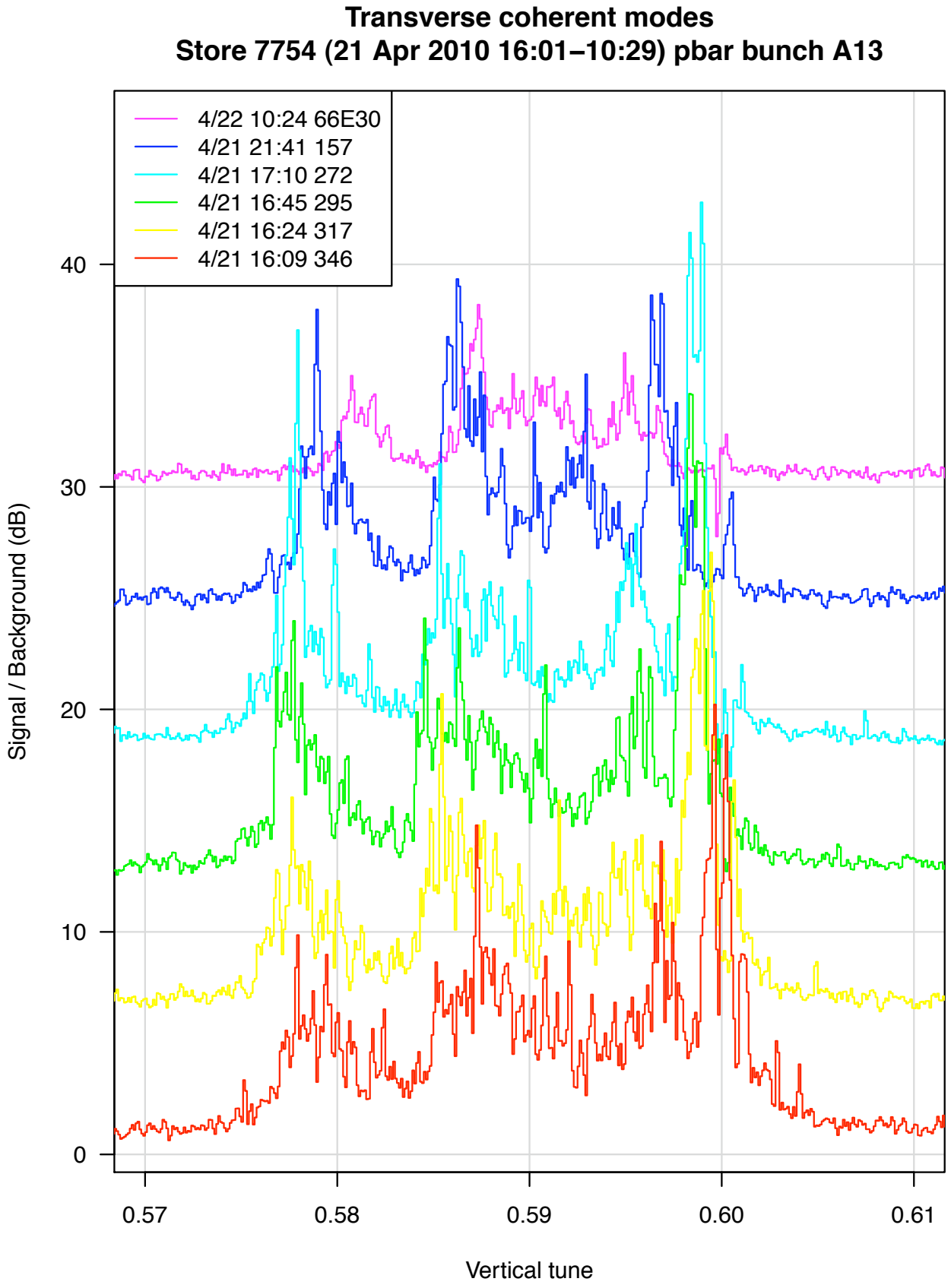

Figure 6: Evolution of vertical coherent modes over the course of a store.

## CONCLUSIONS

We have implemented a system to cleanly detect transverse coherent modes of individual proton and antiproton bunches in the Tevatron. The device is used for measurements of the lattice tune and to independently estimate the beam-beam parameter. We plan to employ in our ongoing studies of compensation of nonlinear beam-beam forces with Gaussian electron lenses.

## ACKNOWLEDGEMENTS

We would like to thank Y. Alexahin, B. Fellenz, G. Saewert, V. Scarpine, and V. Shiltsev (Fermilab) for their help and insights.